\documentclass[journal]{IEEEtran}
\usepackage{cite}
\usepackage{amsmath,amssymb,amsfonts}
\usepackage{graphicx}
\usepackage{booktabs}
\usepackage{url}
\usepackage{xspace}

\newcommand{\eagerAtwo}{24.5}   \newcommand{\eagerAfourteen}{72.4}  \newcommand{\eagerAonefourty}{573.0}
\newcommand{\eagerBonefourty}{448.4} \newcommand{\eagerDonefourty}{392.5}
\newcommand{\eagerBdelta}{22\%} \newcommand{\eagerDdelta}{31\%}
\newcommand{\swarm}{11.0} 
\newcommand{\lazyttfptwo}{16.9} \newcommand{\lazyttfpfourteen}{17.4} \newcommand{\lazyttfponefourty}{17.6}
\newcommand{\lazyreadtwo}{12.0} \newcommand{\lazyreadfourteen}{105.3}
\newcommand{\fusetax}{3.5$\times$}
\newcommand{\liewindow}{196}    \newcommand{\lieeio}{8}
\newcommand{\pmEIOs}{595, 662, 726, 797, and 837}
\newcommand{\pmCorrupt}{188, 209, 231, 252, and 264}
\newcommand{\pmFiles}{280}
\newcommand{\pmOnsetFast}{49}   \newcommand{\pmOnsetSlow}{358}
\newcommand{\socireadyA}{73.2}  \newcommand{\socireadyB}{59.9} \newcommand{\socirx}{15.8}
\newcommand{\rwgzip}{21.9\%}
\newcommand{\rweagerA}{111}     \newcommand{\rwlazy}{12.5--16.7}

\newcommand{\estargz}{eStargz\xspace}
\newcommand{\soci}{SOCI\xspace}

\begin{document}

\title{The Lazy Pod That Lies: Deferred Cost and Failure Semantics of\\
Lazy Container Image Pulling for Model Serving on Kubernetes}

\author{Georgii~Kliukovkin%
\thanks{G. Kliukovkin is an independent researcher, Sunnyvale, CA USA (e-mail: kliukovkin@gmail.com).}}

\maketitle

\begin{abstract}
Lazy container-image pulling promises to eliminate the dominant cost of
starting a model-serving pod by mounting the image immediately and
fetching content on demand. We evaluate this promise for model delivery
on Kubernetes, using KServe with two production lazy-pulling
systems---\estargz/stargz-snapshotter and AWS \soci---against eager
baselines, on artifacts from 2 to 140\,GB including real fp16 weights.
Lazy pulling delivers its headline: cold time-to-first-prediction becomes
size-independent
(\lazyttfptwo--\lazyttfponefourty\,s, versus
\eagerAtwo--\eagerAonefourty\,s eager). But the cost is deferred, not
eliminated: a full read of a 14\,GB model through the lazy mount takes
\lazyreadfourteen\,s, slower than the \eagerAfourteen\,s eager pull it
replaced, and the two systems pay at opposite lifecycle ends (\soci
prefetches nearly the full image before Ready; \estargz defers nearly
everything to first read). More consequentially, we characterize a
failure mode eager pulling structurally cannot exhibit: under sustained
legitimate reads with default configuration, the snapshotter's
node-level cache exhausts its finite volume and already-running pods
begin failing reads of model files. At the earliest
stage of exhaustion, an instrumented serving pod passed every
Kubernetes-visible and application-level check for 196\,s while its
snapshotter was already logging real failures; under heavier pressure,
67--94\% of model files fail, scaling monotonically with residual cache
occupancy. A live pod self-heals if cache space is freed under it, but a
snapshotter-daemon restart under a live pod leaves permanently stale file
handles in a pod still reported Running. We derive placement, monitoring,
and cache-sizing guidance for serving platforms and operators.
\end{abstract}


\section{Introduction}
Serving a machine-learning model on Kubernetes begins with moving its
weights. As models have grown from megabytes to hundreds of gigabytes,
the delivery step---pulling the artifact to the node that will serve
it---has come to dominate pod cold-start time and, through it, the
economics of autoscaling and scale-to-zero. Prior work, including our own
measurements of OCI-based model delivery~\cite{kliukovkin2026coldstart},
shows eager image pulls scaling linearly with size: an eager cold start
that takes \eagerAtwo\,s for a 2\,GB model takes \eagerAonefourty\,s
(nearly ten minutes) at 140\,GB.

Lazy image pulling attacks exactly this. Systems such as
\estargz/stargz-snapshotter~\cite{estargz} and AWS
SOCI~\cite{soci} reformat or index the image so a snapshotter can mount it
immediately and fetch content on demand, making container start time
independent of image size. The technique ships in production systems, and its
appeal for model serving is obvious: our measurements confirm cold
time-to-first-prediction of \lazyttfptwo--\lazyttfponefourty\,s
\emph{regardless of size} from 2 to 140\,GB---roughly 33$\times$ faster
than the single-layer eager control at the large end (22$\times$ vs.\ the
best eager packaging).

The literature on lazy pulling measures start-up acceleration. What it
does not characterize is the other side of the ledger: where the deferred
cost reappears, and what new failure behavior the deferral introduces into
a running serving fleet. Those are precisely the questions an operator
must answer before enabling lazy pulling for production model serving,
and, to our knowledge, no controlled study answers them.

This paper makes four contributions:
\begin{itemize}
\item \textbf{Deferred-cost accounting.} We quantify where lazy
  pulling's saved cost reappears---and where it does not: at 2\,GB the
  deferred full read is \emph{cheaper} than the eager pull it replaced,
  but by 14\,GB it costs \lazyreadfourteen\,s against
  \eagerAfourteen\,s eager, is not network-bound, and carries an
  isolated \fusetax{} lazy-mount overhead even when fully warm
  (Section~\ref{sec:price}).
\item \textbf{Architectural contrast of two production systems.} \soci
  and \estargz move the same bytes at opposite ends of the pod
  lifecycle: \soci transfers $\sim$\socirx\,GB \emph{before} Ready
  (Ready at $\sim$60--74\,s; the first read is then effectively local),
  while \estargz reaches Ready in $\sim$12--17\,s with
  near-zero transfer and pays at first read. Front-loaded versus
  deferred is the real design axis, not ``lazy versus eager.''
\item \textbf{Failure-semantics characterization.} We induce, quantify,
  and dissect the ``lying pod'': a \liewindow-second window of green
  Ready/health/predict signals over a pod whose snapshotter is already
  logging real I/O errors; a monotonic pressure curve
  (67--94\% of files corrupted, the failing read completing
  in \pmOnsetSlow\,s down to \pmOnsetFast\,s as residual cache occupancy
  grows); a recovery matrix including an unrecoverable daemon-restart
  outcome; eager symmetry controls; and a detectability inventory
  (Section~\ref{sec:lie}).
\item \textbf{Operational guidance.} A placement rule by workload read
  fraction, node-wide cache capacity planning, snapshotter telemetry as
  a first-class serving signal, and daemon-restart hygiene
  (Section~\ref{sec:guidance}).
\end{itemize}

\section{Background and Related Work}
\subsection{Container image delivery and lazy pulling}
An OCI container image is a stack of content-addressed layers, each a
compressed tar archive. The conventional (\emph{eager}) pull downloads
every layer and unpacks it into the node's snapshotter storage before the
container starts; pull time therefore scales with image size, and the
container runtime's readiness machinery only proceeds once the complete
filesystem is materialized. Harter et al.\ measured a decade ago that
containers typically read only a small fraction of the image bytes they
pull~\cite{slacker}, and a line of systems work has attacked the resulting
waste: Slacker's lazy NFS-backed images~\cite{slacker}, DADI's block-level
remote image layers~\cite{dadi}, Starlight's delta-oriented
delivery~\cite{starlight}, and FaaSNet's function-oriented image
distribution~\cite{faasnet}.

Several lazy-pulling snapshotters are shipped for mainstream containerd
today---including Nydus, the Dragonfly project's RAFS-based image
service~\cite{nydus}---and we evaluate the two most widely deployed, which
take instructively different approaches. \estargz~\cite{estargz}
re-encodes each layer into a \emph{seekable} tar.gz with a table of
contents, so that stargz-snapshotter---a containerd remote-snapshotter
plugin---can mount the layer immediately over FUSE and fetch individual
chunks on demand with HTTP range requests against the registry; an
optional ``landmark'' mechanism prefetches files expected to be needed at
startup. \soci~\cite{soci} instead leaves the image untouched and builds a
separate index artifact (a ``ztoc'') stored alongside it; its snapshotter
also mounts lazily, but a background fetcher proactively downloads layer
content concurrently with container startup. In both systems the
container's reads traverse a FUSE filesystem served by a node-level
daemon, and fetched content accumulates in a node-level on-disk cache
shared by every lazily mounted image on the node---two facts that are
central to this paper.

\subsection{Model serving on Kubernetes and the delivery problem}
Model-serving platforms materialize a model artifact on a node before a
serving pod can answer its first request. In KServe~\cite{kserve}, a widely
deployed CNCF (incubating) serving platform we use as our measurement
vehicle, the artifact arrives
either through a storage-initializer download from object storage, or as
an OCI image---as a ``modelcar'' sidecar or, since the platform's recent
OCI-native work, as a Kubernetes image volume (KEP-4639~\cite{kep4639})
mounted directly into the inference container. Our earlier study of these
eager delivery paths~\cite{kliukovkin2026coldstart} found cold-start
delivery time scaling linearly to $\sim$40 minutes of re-download for a
140\,GB-class artifact and identified the download-then-unpack structure
of eager OCI pulls as a $\sim$2$\times$ cold-start tax. Model serving is
also the workload where lazy pulling's value proposition is strongest:
autoscaling and scale-to-zero multiply the cold-start cost by every
replica added under load, and model images are precisely the
hundreds-of-gigabytes artifacts for which eager pulls are slowest.

\subsection{What is missing}
The evaluation sections of the lazy-pulling literature measure one side of
the ledger: container start-up latency, occasionally with aggregate
bandwidth savings. None of the systems above characterizes (i) where the
deferred delivery cost reappears in the pod's lifetime, (ii) how two
architecturally different production implementations distribute that cost,
or (iii) what happens to \emph{already-running} pods when the shared
node-level cache that lazy pulling introduces comes under pressure. For an
operator deciding whether to enable lazy pulling under a production
serving fleet, those three questions are the decision; this paper answers
them with controlled measurements.

\section{Methodology}
\subsection{Testbed}
All measurements ran on a two-host rig in one AWS availability zone
(us-east-1): a \emph{registry host} serving an OCI registry
(\texttt{registry:2}, plain HTTP) and an S3-compatible object store
(MinIO), and a \emph{node host} running a single-node Kubernetes cluster
(kind, \texttt{kindest/node} v1.36.1; containerd 2.3.1 with the
\texttt{ImageVolume} feature enabled) and KServe built from source. Both
hosts are i4i.2xlarge instances (8 vCPU, 64\,GB RAM, 1.9\,TB local NVMe
instance store), so that download (network) and unpack (disk) are
genuinely independent resources---a deliberate correction to an earlier
single-host pilot in which registry and node shared one disk.
Lazy-pulling components: containerd-stargz-grpc v0.18.2 and
soci-snapshotter v0.15.0, each wired in as a containerd
\texttt{proxy\_plugins} remote snapshotter.

The results come from two measurement campaigns on same-model hardware:
a performance campaign (C1: the eager/lazy delivery matrix and full-read
costs) and an instrumented failure-characterization campaign (C2)
designed after an adversarial review of the first. Two provisioning
details changed between them, both in C2's favor: C2 moved the bounded
snapshotter cache from a loopback file to a real NVMe partition, and
added per-rep NIC byte accounting; every number below is labeled with
its campaign where the distinction matters. Each campaign was calibrated
before any benchmark rep and re-calibrated at the end, with no meaningful
drift: node-disk sequential throughput $\sim$1054--1060\,MiB/s write and
$\sim$1341--1367\,MiB/s read (fio, 1-minute), inter-host bandwidth
4.96--4.97\,Gbit/s single-stream and 11.9\,Gbit/s over eight parallel
streams (iperf3), and $\sim$494\,MiB/s ($\approx$518\,MB/s) for a real 14\,GB blob
over the registry's HTTP API. For the bounded-cache failure experiments (C2), the
snapshotter cache lived on a dedicated 92\,GB \emph{real NVMe partition}
(not a loopback file), deliberately smaller than the largest image's
$\sim$150\,GB of content.

\subsection{Artifacts and workloads}
Model images combine a small real scikit-learn model with incompressible
random ballast at three size tiers---2, 14, and 140\,GB---packaged as
(A) one gzip layer, (B) eight gzip layers, and (D) eight zstd layers;
\estargz images are conversions of (B) via \texttt{ctr-remote}. Ballast is
incompressible by construction, mimicking the near-random statistics of
trained fp16 weights while keeping image construction controlled. To
bound the realism gap, a separate arm packages actual open weights
(Mistral-7B-v0.1, fp16 safetensors, 14.48\,GB raw, Apache-2.0): under
gzip these compress by \rwgzip---inside the 5--30\% band expected for
real weights---so we report both ballast and real-weight results where
they differ.

Time-to-first-prediction (TTFP) is measured from \texttt{kubectl apply}
to the first successful HTTP 200 from the model's \texttt{:predict}
endpoint with a real inference payload---not a liveness or readiness
endpoint. A per-phase decomposition (Section~\ref{sec:promise}) confirms
first-predict-200 lands within 0.08--0.18\,s of Ready, so TTFP is not
inflated by post-Ready effects. Sustained-read experiments read every
file of the mounted model directory (\texttt{find~|~cat}), an upper bound
for a server that touches all weight bytes. For failure
characterization we additionally built a \emph{datapath-coupled}
predictor: a container whose \texttt{:predict} handler reads three model
files per request, chosen by a client-supplied seed, and returns SHA-256
fragments of the bytes read. Our sampler drives it with a fixed seed, so
the probe reads a fixed, reproducible 3-of-\pmFiles{} file set (about
1\% of the ballast) on every request and fails loudly (HTTP 503) if any
read fails; responses are checked in-loop for structural integrity
(well-formed, error-free), with the fixed seed additionally enabling
offline comparison of the returned hashes against a healthy-pod
baseline (predictor and sampler source are in the artifact repository).

\subsection{Measurement hygiene}
Every cold rep runs under an enforced cold protocol: CRI image removal
and prune, snapshotter cache directories wiped and verified empty, and
page caches dropped on both hosts. Coldness is then \emph{verified}, not
assumed: per-rep NIC byte counters on the node must show at least the
image's content size transferred, or the rep is flagged. This check
caught a real contamination bug during the study---for \soci, two cache
layers outside the snapshotter's own directory tree (containerd's
content-addressed store and \soci's persistent blob cache) survive
\texttt{crictl rmi -{}-prune} entirely, so a na\"ive protocol measures
warm-cache behavior mislabeled as cold; the contaminated dataset is
preserved in the artifact release but excluded from all reported numbers.
Each rep persists an evidence bundle (pod and probe specs, events,
snapshotter and runtime configs, image digests, logs, metrics scrapes,
disk and NIC samples). We report medians with full ranges and publish every raw rep. All
excluded data is preserved and flagged in the artifact rather than
deleted: one invalidated first attempt at the pressure matrix, the
contaminated \soci performance dataset above, four failed attempts at
an amply cached 140\,GB lazy read, one cache-contaminated \estargz
read rep, and two \soci timeout reps traced to a harness race.

\section{The Promise: Size-Independent Cold Start}
\label{sec:promise}
\begin{table}[t]
\caption{Cold time-to-first-prediction (s; medians, campaign C1).
N=5 per cell, except eager 140\,GB (N=3) and the s3 calibration row
(N=1 per size).}
\centering
\begin{tabular}{lccc}
\toprule
Scheme & 2\,GB & 14\,GB & 140\,GB \\
\midrule
Eager, 1-layer gzip & \eagerAtwo & \eagerAfourteen & \eagerAonefourty \\
Eager, 8-layer gzip & 21.4 & 57.8 & \eagerBonefourty \\
Eager, 8-layer zstd & 20.6 & 52.7 & \eagerDonefourty \\
Object storage (s3) & 22.8 & 60.9 & 461.2 \\
\estargz (lazy) & \lazyttfptwo & \lazyttfpfourteen & \lazyttfponefourty \\
\bottomrule
\end{tabular}
\label{tab:ttfp}
\end{table}

Table~\ref{tab:ttfp} shows the delivery matrix. Eager cold starts scale
linearly with size regardless of packaging: multi-layer images reduce the
140\,GB cold start by \eagerBdelta{} (layer-level pipelining of download
and unpack) and zstd compression by \eagerDdelta{} in total, consistent
with our earlier eager-path findings~\cite{kliukovkin2026coldstart}, but
the linear shape is untouched---at 140\,GB the best eager variant still
needs six and a half minutes. \estargz replaces the shape itself: cold
TTFP is \lazyttfptwo, \lazyttfpfourteen, and \lazyttfponefourty\,s at 2,
14, and 140\,GB (N=5 each; full spread across all fifteen reps
16.4--18.0\,s)---practically size-independent, with median growth of
0.7\,s across a 70$\times$ size range: a 32.6$\times$ speedup over the
single-layer eager control at 140\,GB and 22$\times$ over the best
eager packaging. Warm eager starts (image resident) take
$\sim$\swarm\,s at every size (one verified-warm rep per size; the
preceding warm-up rep repopulates the cache), so lazy pulling's cold
start is within $\sim$7\,s of an eager warm start. The kubelet-reported ``pull''
for these starts is 59\,ms--1.4\,s: a manifest fetch, not a data
transfer.

\textbf{Where the 17 seconds go.} A per-phase decomposition (raw-Pod
deployment path, N=5) shows scheduling, pull, and container start all
completing within $\sim$1\,s of apply; the entire remaining budget---
Ready at median 11.8\,s---sits between container start and readiness:
probe polling plus the lazy first read of the (small) model file itself.
First predict-200 follows Ready within 0.08--0.18\,s
(Table~\ref{tab:decomp}).

\begin{table}[t]
\caption{TTFP decomposition, \estargz 14\,GB, raw-Pod path (C2, N=5).
Scheduling, image pull, and container start each complete within
$\sim$1\,s of apply (pull is a manifest-only fetch; pod-event
timestamps are second-granularity, so per-phase offsets of $\pm$0.5\,s
occur); the budget sits in container-start$\rightarrow$Ready.}
\centering
\begin{tabular}{ccc}
\toprule
Rep & Ready (s) & First predict-200 (s) \\
\midrule
1 & 12.84 & 12.93 \\
2 & 11.83 & 11.92 \\
3 & 10.81 & 10.99 \\
4 & 11.78 & 11.86 \\
5 & 11.79 & 11.87 \\
\midrule
median & 11.79 & 11.87 \\
\bottomrule
\end{tabular}
\label{tab:decomp}
\end{table} The decomposition
(11.8\,s median, raw-Pod path, C2) sits $\sim$5.6\,s below the
InferenceService-managed figures of Table~\ref{tab:ttfp} (C1); the gap
is consistent with control-plane reconciliation overhead in the managed
path, but the two figures come from different campaigns and deployment
paths and we did not isolate the overhead on a single rig
(Section~\ref{sec:threats}).
Both are reported as measured.

\textbf{The speed is not prefetch curation.} \estargz supports a
``landmark'' prefetch list, and a natural suspicion is that fast Ready
reflects a lucky or curated landmark set. A per-layer audit of a fresh
deployment shows \texttt{prefetch\_size=0} for every layer: our converted
images prefetch nothing, and fast Ready is achieved by genuinely deferring
all data movement. This makes the result a floor, not a best case.

\textbf{Real weights change nothing here.} On Mistral-7B fp16 weights
(C2), \estargz TTFP is \rwlazy\,s (N=3) versus a median of
$\sim$\rweagerA\,s for the eager single-layer control (N=5) at the same
14\,GB-class size---the lazy TTFP advantage is insensitive to whether
the payload is synthetic ballast or real model weights. Real weights do,
however, change the compression picture that incompressible ballast
hides (Table~\ref{tab:compress}): gzip recovers \rwgzip{} on real fp16
weights, squarely inside the 5--30\% band expected for trained
parameters, while the zstd figure is confounded by the image builder's
default (speed-oriented) compression level and is flagged rather than
claimed.

\begin{table}[t]
\caption{Compressed size of the Mistral-7B fp16 image variants (C2;
raw payload 14.48\,GB).}
\centering
\begin{tabular}{lcc}
\toprule
Variant & Compressor & Compressed / raw \\
\midrule
A (1 layer) & gzip & 78.1\% ($-$21.9\%) \\
B (8 layers) & gzip & 78.1\% ($-$21.9\%) \\
D (8 layers) & zstd (builder default) & 100.0\% (confounded) \\
\estargz (from B) & gzip, \estargz-framed & $\sim$78\% \\
\bottomrule
\end{tabular}
\label{tab:compress}
\end{table}

\section{The Price: Deferred Cost}
\label{sec:price}
\subsection{Full-read cost and the FUSE tax}
The deferred bytes come due the first time the workload actually reads
them---but which side of the ledger wins depends on size. At 2\,GB the
deferred read costs \lazyreadtwo\,s against a \eagerAtwo\,s eager
pull: deferral wins outright, since fixed per-chunk overheads do not yet
dominate. By 14\,GB the sign flips: the cold full read takes
\lazyreadfourteen\,s (median, N=3, C1; reproduced in C2 at
102.0/104.9\,s)---\emph{slower than the \eagerAfourteen\,s eager
pull-and-unpack it replaced}. In C2, per-rep NIC counters confirm the
14\,GB reads genuinely moved $\sim$15\,GB over the wire, and also show
the read is not network-bound: sustained throughput is
$\sim$143\,MB/s against a measured $\sim$518\,MB/s achieved by a plain
HTTP blob download over the same registry path. The gap is the lazy data
path itself---per-chunk HTTP round trips, per-chunk decompression, and
the FUSE/snapshotter userspace traversal. We isolate that last component
directly: re-reading fully-resident content through the lazy mount takes
6.37\,s (N=2) where a direct read of equivalent eager-mounted content
takes 1.81\,s (N=1)---a \fusetax{} lazy-mount overhead that remains
after all fetching and decompression is done. Warm reads drop to
6.3--7.7\,s at 14\,GB across both campaigns (and to sub-second at
2\,GB once fully resident; the first warm pass still pays a residual
$\sim$10\,s), so the penalty is a first-touch cost, not a permanent
one.
We could not obtain a clean 140\,GB full read with an amply sized cache
(four attempts failed for environment reasons documented in the artifact
release), so the deferred-cost curve is quantified at 2 and 14\,GB and
qualitatively confirmed at 140\,GB.

\subsection{Two architectures, one bill: front-loaded vs.\ deferred}
\soci distributes the same cost differently. With coldness enforced and
verified (Section~III-C), a 14\,GB \soci deployment reaches Ready in
\socireadyA\,s (single-layer) or \socireadyB\,s (8-layer, N=2 each,
ranges $\le$2.1\,s)---and the NIC shows $\sim$\socirx\,GB transferred
\emph{before Ready}, at or above the image's full content size: \soci
pre-pulls essentially the whole image during the startup window,
consistent with its background-fetcher design. A post-Ready full read is
then effectively local---near-zero additional network transfer; the
$\sim$1.9\,s read times we observed in reps flagged as cache-warm are a
consistent lower bound, though we did not obtain a separately
clean-protocol read measurement. \estargz is the mirror
image: Ready in $\sim$12--17\,s with near-zero transfer, then
$\sim$102--105\,s of fetching on first read. Both systems move the same
bytes; they differ in \emph{when}---\soci front-loads the cost where
operators already watch (startup), \estargz defers it to a place nobody
instruments (first data access). Neither is free, and the choice between
them is a choice about where in the pod lifecycle the bill lands. The
8-layer \soci image reaches Ready $\sim$18\% faster than the single-layer
one, consistent with per-layer parallelism in the background fetcher; the
packaging guidance from the eager world carries over.

\section{The Lie: Failure Semantics Under Cache Pressure}
\label{sec:lie}

\subsection{A threat-free failure model}
Everything in this section happens with no attacker, no fault injection
into the lazy data path, and no misconfiguration: the only ingredients
are default snapshotter configuration and sustained \emph{legitimate}
reads. The ingredients are structural. Every lazily-fetched chunk on a node is
written into a shared node-level cache directory tree: stargz-snapshotter
v0.18.2 maintains two on-disk trees under one root---an HTTP blob-chunk
cache (\texttt{httpcache}) and a filesystem cache
(\texttt{fscache})---both of which grew throughout our inductions, with
the ENOSPC failures below occurring at the blob-cache write path. By
inspection of the v0.18.2 source (not by measurement), no shipped
configuration bounds this tree's on-disk size: the available knobs
(\texttt{max\_lru\_cache\_entry}, \texttt{max\_cache\_fds}) bound
in-memory LRU entries and open file descriptors, not bytes on disk, and
cache directories are cleaned only per-layer when an image is removed
(upstream issue \#1213~\cite{stargz1213} requests explicit cache
management). ``Run with defaults'' therefore means ``run with an
unbounded, shared, node-level cache''---and when its volume fills, the
failure lands not at pull time but in the read path of pods that are
already Ready and serving. We characterize this failure
phenomenologically; the internal path by which a failed cache write
becomes a persistent read error in v0.18.2 was not root-caused (planned
configuration micro-experiments were abandoned for environment-stability
reasons; see the artifact).

\subsection{The lying-pod timeline}
\label{sec:lying}
\begin{figure}[t]
\centering
\includegraphics[width=\columnwidth]{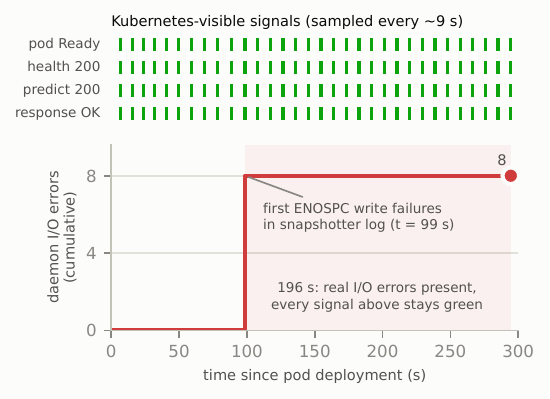}
\caption{The lying-pod timeline (C2). Top: every Kubernetes-visible
signal (Ready condition, health endpoint, predict endpoint with
structural response check) sampled every $\sim$9\,s over a
140\,GB-class lazy pod under cache pressure---green at every sample.
Bottom: the snapshotter daemon's cumulative I/O error count over the
same window; \lieeio{} real ENOSPC write failures appear at
$t{=}99$\,s and persist for the remaining \liewindow{}\,s (the end of
the experiment window) while all signals above stay green.}
\label{fig:lying}
\end{figure}

We deployed the datapath-coupled predictor (Section~III-B) on a
140\,GB-class \estargz image against a cache already under pressure from
earlier legitimate reads of smaller images, and sampled every
$\sim$9\,s: pod Ready condition, restart count, health endpoint,
predict endpoint (status, latency, and structural integrity of the
response), the daemon's cumulative I/O error count, and cache occupancy.
At $t{=}99$\,s the snapshotter log records its first real ENOSPC write
failures (\lieeio{} errors). For the remaining \liewindow{} consecutive
seconds---every sample until the experiment window ended, so
\liewindow{}\,s is a lower bound---the pod reported:
\begin{center}
\texttt{ready=True restarts=0 health=200 predict=200}
\end{center}
with every predict response well-formed and error-free.
Figure~\ref{fig:lying} shows the timeline. Two honest qualifications
sharpen, rather than weaken, what this window shows. First, the probe's
fixed 3-file set covers $\sim$1\% of the ballast; even if all \lieeio{}
affected files had been client-unreadable, a probe of that coverage
misses them with probability $\approx$0.92---and ours did. Second,
whether those first \lieeio{} cache-write failures made any file
client-unreadable at all is not established by this experiment (a failed
cache write can degrade to a refetch); client-visible corruption under
the same mechanism is established directly by the pressure matrix below,
where 67--94\% of files fail. The window therefore demonstrates the
observability gap precisely at the stage an operator most needs a
signal: the earliest phase of cache exhaustion, when the daemon is
already failing, every Kubernetes-visible and client-visible indicator
is still green, and the node is minutes away from mass corruption. A
datapath-coupled probe is not a cure but a dice roll whose odds are set
by probe coverage versus corruption fraction; at the fractions induced
below, any probe notices---by then the damage is done.

\subsection{Failure scales with residual cache occupancy}
\begin{table}[t]
\caption{Pressure matrix (C2): 140\,GB-class full read against a
92\,GB cache partition pre-filled with other images' chunks (N=2 per
level, independent inductions from verified-clean state). Nominal
pre-fill levels correspond to measured occupancies of 1, 28, 53--54, 80,
and 95\% at read start. ``Completion'' is the elapsed time of the
failing full read; time-to-first-error was not separately captured.}
\centering
\begin{tabular}{cccc}
\toprule
Pre-fill & I/O errors & Corrupted (of \pmFiles) & Completion (s) \\
\midrule
0\%  & 594 / 596 & 188 / 187 & 357 / 359 \\
25\% & --- / 662 & --- / 209 & 901$^{\dagger}$ / 277 \\
50\% & 723 / 728 & 231 / 231 & 172 / 174 \\
75\% & 796 / 798 & 252 / 251 & 95 / 96 \\
90\% & 837 / 837 & 264 / 264 & 51 / 47 \\
\bottomrule
\end{tabular}
\label{tab:pressure}
\\[2pt]{\footnotesize $^{\dagger}$Rep 1 at 25\% is the run's one
non-monotonic outlier: the pod never reached Ready within 900\,s and no
container existed to inspect---a \emph{heavier} failure than either
neighboring level, showing failure character varies even where severity
trends cleanly.}
\end{table}

\begin{figure}[t]
\centering
\includegraphics[width=\columnwidth]{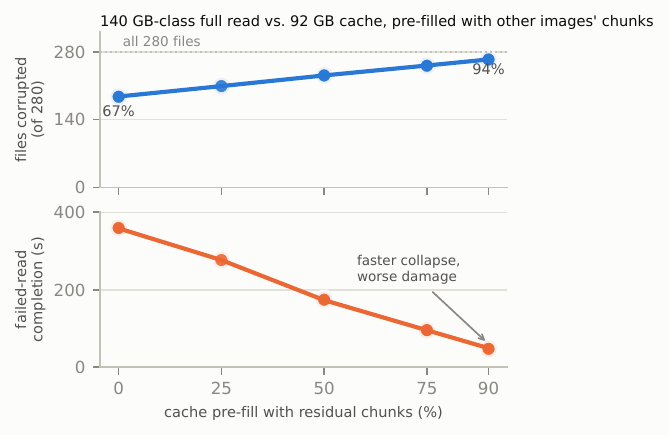}
\caption{Pressure response (C2; data of Table~\ref{tab:pressure}; dots
are individual inductions, lines connect means). Damage grows, and the
failing read collapses faster, monotonically with residual cache
occupancy from other images' chunks.}
\label{fig:pressure}
\end{figure}

Table~\ref{tab:pressure} and Figure~\ref{fig:pressure} make the failure
quantitative. As residual
occupancy from other images' chunks rises from 0\% to 90\% (nominal),
damage grows monotonically---\pmEIOs{} I/O errors (means); \pmCorrupt{}
of \pmFiles{} files corrupted (67--94\%)---and the failing
read completes ever faster, from \pmOnsetSlow\,s down to
\pmOnsetFast\,s (means): at high occupancy most reads fail nearly
instantly. We did not capture time-to-first-error within a read, so we
make no claim about when errors begin, only how severity and total
failure duration scale. The operative variable is
\emph{cumulative, cross-image} cache pressure: the risk is a property of
everything the node has lazily touched, not of any single image. We are
deliberately careful at the left edge of the table: across our two
campaigns, nominally clean-cache full reads of the oversized image both
succeeded (98\,GB cache, C1 rig; two instrumented reruns completing all
280 files with zero errors, preserved in the artifact) and failed
(92\,GB partition, Table~\ref{tab:pressure})---the zero-residual boundary is
margin-sensitive, and we claim the monotonic curve, not a deterministic
outcome at either end of it.

\subsection{Recovery matrix}
\label{sec:recovery}
What can an operator do once a pod is in this state? We tested the
remediation ladder on one live induced failure (single trial throughout
this subsection), in order; Table~\ref{tab:recovery} summarizes.

\begin{table}[t]
\caption{Recovery matrix (C2, one instrumented induced failure).}
\centering
\begin{tabular}{p{0.44\columnwidth}p{0.44\columnwidth}}
\toprule
Action & Outcome \\
\midrule
Re-read failed pod, 3$\times$ & corrupted-file set identical each time
(224 files) \\
Free cache space under the live pod (no restarts) & live mount heals;
previously failing file reads cleanly at $\sim$118\,MB/s \\
Replace pod after space freed & replacement pod clean \\
Restart snapshotter daemon under the live pod & permanent
\texttt{ESTALE} on reads through the mount; pod still
\texttt{1/1 Running} \\
\bottomrule
\end{tabular}
\label{tab:recovery}
\end{table} (1)~The corrupted-file set is \emph{stable}:
three consecutive re-reads of the failed pod yielded an identical list of
224 broken files---the failure is not transient flakiness within a fixed
pressure state.
(2)~Freeing cache space \emph{under the live pod} (deleting other
images' cached chunks; no restart of anything) heals it: a previously
failing file then read successfully through the same FUSE mount at
$\sim$118\,MB/s. The failure is fully reversible in place---if the
operator knows to act. (3)~Replacing the pod after space is freed is
likewise clean. (4)~But restarting the \emph{snapshotter daemon} under a
live pod---the reflexive first move for a node agent misbehaving---makes
things permanently worse: subsequent reads through the pod's mount fail
with \texttt{ESTALE} (``Stale file handle''), unrecoverable for the
life of the pod, while \texttt{kubectl} continues to report the pod
\texttt{1/1 Running}. This is an instance of the general hazard class of
FUSE-backed mounts orphaned by their userspace daemon's death, observed
here in one trial on stargz-snapshotter; \soci ships a FUSE-manager
mode designed to preserve mounts across daemon restarts, which we did
not test. Note also the errno progression the workload observes:
under cache pressure, affected reads fail with I/O errors
(\texttt{EIO}); after a daemon restart, reads through the orphaned
mount fail with \texttt{ESTALE}. Neither errno points at a disk-space root cause on
another filesystem the workload has never heard of.

\subsection{Eager symmetry controls}
Two controls pin down what is and is not lazy-specific. First,
\emph{eager under the same disk exhaustion fails loud and early}: with
free space below the image size, an eager 140\,GB deployment did not
reach Ready within a 900\,s observation window (single trial)---the
failure surfaces pre-Ready as image-pull backoff in pod events, exactly
where operators and autoscalers already look; and because the exhausted
filesystem in the eager case is kubelet's own image filesystem, the
condition additionally feeds kubelet's disk-pressure and garbage-collection
machinery~\cite{kubeleteviction}.
Second, \emph{Kubernetes' blindness to data corruption is not
lazy-specific}: when we manually corrupted a model file under a healthy
\emph{eager} pod, it too stayed Ready with predict returning 200. The
distinction, and the paper's novelty claim, is therefore precise: lazy
pulling does not create the observability gap---readiness probes never
certified data integrity---but it is the first mainstream delivery
mechanism that \emph{self-inflicts} post-Ready data-path corruption
under default configuration, with zero external action, as a byproduct
of legitimate load. Eager images do not corrupt themselves; something
has to reach in and break them.

\subsection{Detectability}
\label{sec:detect}
``Silent'' is a claim about monitoring, so we inventoried every signal
during the inductions; Table~\ref{tab:detect} summarizes.

\begin{table}[t]
\caption{Detectability inventory during induced cache-exhaustion
failures (C2 inductions; the snapshotter log-line count is from C1, as
labeled).}
\centering
\begin{tabular}{p{0.42\columnwidth}p{0.46\columnwidth}}
\toprule
Signal & During failure \\
\midrule
Pod Ready condition & \texttt{True} throughout \\
Container restarts & 0 \\
Pod events & none \\
Node \texttt{DiskPressure} & \texttt{False} throughout (cache volume
outside kubelet accounting) \\
Node \texttt{stats/summary} fs & no anomaly \\
Kernel log (dmesg) & no FUSE trace \\
Snapshotter Prometheus endpoint & moves; snapshotter-specific, not
scraped by default stacks \\
Snapshotter log & 3{,}415 ENOSPC lines (C1)---loud, but untailed \\
\bottomrule
\end{tabular}
\label{tab:detect}
\end{table} Pod-level: Ready \texttt{True}, restarts 0, no
events. Node-level: \texttt{DiskPressure} \texttt{False} throughout
every induction---the dedicated cache volume is invisible to kubelet's
nodefs/imagefs accounting (and a dedicated volume is a recommended
deployment pattern for exactly these caches, so this is not an artifact
of our setup so much as a property of it); node \texttt{stats/summary}
shows no anomaly; kernel logs show no FUSE-level trace. The places the
failure \emph{is} visible are all snapshotter-specific: the daemon's own
log (3{,}415 ENOSPC lines during C1's inductions---
loud, but a log no standard stack tails) and its Prometheus endpoint,
which no default KServe/Kubernetes monitoring pipeline scrapes. The
accurate statement is therefore not that the failure is silent, but that
it is \emph{invisible to every pod- and node-level Kubernetes signal,
and observable only through snapshotter-specific telemetry that serving
stacks do not collect by default}. Section~\ref{sec:guidance} turns that
into concrete monitoring guidance.

\section{Discussion: Operational Guidance}
\label{sec:guidance}
Our measurements condense into a placement rule and three operational
practices.

\textbf{Place by read fraction.} The variable that decides whether lazy
pulling helps is the fraction of image bytes the workload will actually
touch. Bounded-fraction workloads---adapter serving, sharded or
partially-loaded models, catalogs where any one replica reads a slice---
get the full cold-start win and defer costs they may never pay; and
below the crossover size (single-digit GB in our setup), even full-read
workloads come out ahead. Full-read workloads---dense weights loaded in their entirety at or
soon after startup---should either stay eager (multi-layer zstd
packaging already buys \eagerDdelta{} there) or adopt lazy pulling with
open eyes: the full-read bill arrives at first touch, costs more than
the eager pull it replaced (\lazyreadfourteen{} vs.\
\eagerAfourteen\,s at 14\,GB), and lands on the data path. \soci's
front-loaded architecture is a defensible middle ground for full-read
workloads: Ready arrives later but \emph{honestly}---the cost is paid
where operators already instrument---at the price of losing
size-independent startup.

\textbf{Size the cache for the node, not the image.} The failure
variable in Section~\ref{sec:lie} is cumulative cross-image occupancy.
Cache capacity planning must therefore budget for the sum of all
lazily touched bytes across every image and pod the node will host
between cleanups---not for the largest single image---and, absent
upstream size-bounding (issue \#1213~\cite{stargz1213}), pair the budget with an explicit
hygiene mechanism: image removal is currently the only shipped operation
that reclaims cache space.

\textbf{Monitor the snapshotter, not the pod.} Pod health, restarts,
events, and node \texttt{DiskPressure} all stay green through the
failure (Section~\ref{sec:detect}). The signals that do move are the
snapshotter's own: its Prometheus endpoint and its log, plus plain
occupancy of the cache volume. Any production deployment of lazy pulling
should scrape snapshotter telemetry and alert on cache-volume occupancy
as first-class serving signals.

\textbf{Treat daemon restarts as pod-killing events.} The
\texttt{ESTALE} result in Section~\ref{sec:recovery} (one trial,
stargz-snapshotter) argues for operationally coupling a
snapshotter-daemon restart to draining or replacing every pod with a
lazy mount on that node, unless the deployment uses a mount-preserving
mechanism such as \soci's FUSE manager: restarting the daemon ``under''
live pods converted a recoverable condition into a permanent one. No
default signal surfaces that state---though unlike the partial-corruption
dice roll of Section~\ref{sec:lying}, a liveness probe that reads model
data would catch it immediately, since after the restart \emph{every}
read fails.

\section{Threats to Validity}
\label{sec:threats}
\emph{Environment.} All measurements use a single-node kind cluster
(Kubernetes-in-containers) on one hardware generation; absolute numbers
will shift on multi-node production clusters, though the relative
comparisons and the failure mechanism---which is a property of the
snapshotter's node-local cache, not of the control plane---should carry.
\emph{Systems and versions.} Findings are specific to
stargz-snapshotter v0.18.2 and soci-snapshotter v0.15.0; the
cache-management gap is checkable against upstream source and may be
fixed in future releases---our claims are versioned, not eternal.
\emph{Workload realism.} Most arms use incompressible synthetic ballast;
the real-weights arm bounds the gap for TTFP and compression but was not
repeated for the failure experiments. The zstd comparison on real
weights is confounded by the image builder's default compression level
and is flagged, not claimed. \emph{SOCI failure parity.} We attempted
and could not cleanly reproduce the cache-exhaustion failure on \soci
within our time budget---persistent local caching resisted isolation.
We therefore claim the failure mode for \estargz only and explicitly
do \emph{not} claim \soci is immune; its proactive background fetcher
gives it the same unbounded-write ingredient. \emph{Boundary
sensitivity.} The clean-cache (zero-residual) outcome differed across
our two campaigns (Section~\ref{sec:lie}); we accordingly claim the
monotonic pressure curve, not behavior at the boundary.
\emph{Single-trial results.} The lying-pod timeline, the recovery
ladder (including the \texttt{ESTALE} outcome), and the eager-ENOSPC
control are each one instrumented trial; the pressure matrix (N=2 per
level; N=1 usable at the 25\% level) is the replicated core. \emph{Registry transport.} Our registry
serves plain HTTP inside one security group; TLS would add per-range-request
overhead that our per-chunk fetch costs do not include, so the lazy
read-path numbers are, if anything, favorable to lazy pulling.
\emph{Cross-campaign comparisons.} The InferenceService-vs-raw-Pod TTFP
gap and the 14\,GB full-read reproduction span the two campaigns and are
labeled as such where quoted; the snapshotter-configuration claims in
Section~\ref{sec:lie} rest on source inspection of v0.18.2, not on
measurement.

\section{Conclusion}
Lazy image pulling keeps its headline promise for model serving: cold
start becomes independent of model size, and on real weights as well as
synthetic ones. But the promise is a deferral, not a discount. The bytes arrive
later; above a small crossover size they cost more when they do; and---in
the deferred architecture---they arrive through a node-level cache that,
under shipped defaults, no configuration bounds and (on the dedicated
cache volumes such deployments use) no Kubernetes signal watches. The
result is a failure class new to mainstream container delivery: pods
that corrupt their own data paths under legitimate load while every
health indicator stays green. None of this argues against lazy pulling;
it argues for deploying it as what it is---a scheduling decision about
\emph{when} and \emph{where} delivery cost and delivery risk land---with
cache capacity planned node-wide, snapshotter telemetry promoted to a
first-class serving signal, and readiness understood as a statement
about processes, not data. Our benchmark harness, raw measurements, and
evidence bundles are open source at
\url{https://github.com/kliukovkin/lazy-model-delivery-study}.

\section*{Acknowledgment}
The author discloses that AI-based assistants were used in developing
the benchmark tooling and analysis scripts and in editing the
manuscript; the study design, the experiments, validation of all
results, and responsibility for the content rest with the author.

\begin{IEEEbiographynophoto}{Georgii Kliukovkin}
is an independent researcher and software engineer specializing in the
efficiency and reliability of large-scale retrieval, ranking, and
model-serving systems. He is a contributor to KServe, the CNCF
model-inference platform, where he authored the platform's OCI-native
model delivery paths, and his current research focuses on the delivery
and integrity of large model artifacts in production serving
infrastructure.
\end{IEEEbiographynophoto}


\begin{thebibliography}{12}
\bibitem{kliukovkin2026coldstart} G. Kliukovkin, ``Cold-start model delivery in Kubernetes
inference serving: an empirical study of OCI-based distribution and its integrity,''
arXiv:2607.16596, 2026.
\bibitem{estargz} containerd project, ``stargz-snapshotter: fast container image
distribution plugin with lazy pulling,''
\url{https://github.com/containerd/stargz-snapshotter}, v0.18.2, 2026.
\bibitem{soci} AWS Labs, ``soci-snapshotter: a containerd snapshotter plugin for
lazy loading with the Seekable OCI (SOCI) format,''
\url{https://github.com/awslabs/soci-snapshotter}, v0.15.0, 2026.
\bibitem{nydus} Dragonfly project, ``Nydus: the Dragonfly image service,''
\url{https://github.com/dragonflyoss/nydus}, 2026.
\bibitem{slacker} T. Harter, B. Salmon, R. Liu, A. C. Arpaci-Dusseau, and
R. H. Arpaci-Dusseau, ``Slacker: fast distribution with lazy Docker containers,''
in \emph{Proc. USENIX FAST}, 2016.
\bibitem{dadi} H. Li, Y. Yuan, R. Du, K. Ma, L. Liu, and W. Hsu, ``DADI:
block-level image service for agile and elastic application deployment,''
in \emph{Proc. USENIX ATC}, 2020.
\bibitem{starlight} J. L. Chen, D. Liaqat, M. Gabel, and E. de Lara, ``Starlight:
fast container provisioning on the edge and over the WAN,'' in \emph{Proc. USENIX
NSDI}, 2022, pp. 35--50.
\bibitem{faasnet} A. Wang, S. Chang, H. Tian, H. Wang, H. Yang, H. Li, R. Du, and
Y. Cheng, ``FaaSNet: scalable and fast provisioning of custom serverless container
runtimes at Alibaba Cloud Function Compute,'' in \emph{Proc. USENIX ATC}, 2021.
\bibitem{kep4639} Kubernetes Enhancement Proposal 4639, ``OCI VolumeSource,''
\url{https://github.com/kubernetes/enhancements/issues/4639}, 2024.
\bibitem{kserve} KServe Project, ``KServe: standardized distributed generative and
predictive AI inference platform on Kubernetes,''
\url{https://github.com/kserve/kserve}, CNCF incubating project, 2026.
\bibitem{stargz1213} containerd/stargz-snapshotter issue \#1213, ``Clear cache
directory / cache management,''
\url{https://github.com/containerd/stargz-snapshotter/issues/1213}, accessed 2026.
\bibitem{kubeleteviction} Kubernetes Documentation, ``Node-pressure Eviction,''
\url{https://kubernetes.io/docs/concepts/scheduling-eviction/node-pressure-eviction/}, 2026.
\end{thebibliography}
\end{document}